# Design and fabrication of a 2.5T superconducting dipole prototype based on tilted solenoids

CHEN Yu-Quan(陈玉泉)[1,2;1)] MA Li-Zhen(马力祯)[1] WU-Wei(吴巍)[1] WU Bei-Min(吴北民)[1]
YANG Tong-Jun(杨通军)[1] LIANG Yu(梁羽)[1,2]

[1] Institute of Modern Physics, Chinese Academy of Sciences, Lanzhou 730000, China
[2] University of Chinese Academy of Sciences, Beijing 100049, China

**Abstract:** This paper describes a new design of superconducting dipole magnet prototype by the use of tilted solenoids. The magnet prototype, which consists of four layers of superimposed tilted solenoids with operating current of 3708 A, will produce a 2.5 T magnetic field in an aperture of 50 mm diameter. The detailed magnetic field design by using two kinds of software is presented. And their results show a good agree in the magnetic fields. So far we have accomplished the prototype construction and expect a cryogenic test. The process of the magnet fabrication is also reported in detail.

PACS：41.85.Lc；84.71.Ba

## 1 Introduction

In the last decade several superconducting dipole magnets that based on the concept of tilted solenoids have been constructed [1-6]. This type of winding configuration, which superposes two concentric and oppositely tilted solenoids with an angle of α to the central axis (see Fig.1), can produce a perfectly pure dipole field in a large fraction of the coil aperture. These tilted solenoids must be assembled in pairs, that is, the layer of tilted solenoids is even, so that the solenoid component of the field can be canceled while only the dipole field component left. This novel approach has great potential in application to accelerator magnets, gantry and NMR for the advantage that the field is extremely uniform and the effective filed region is very large without any optimization and it is easy to be modularized. However, cancelling one field component leads to reduced efficiency for the field excitation of this magnet which means more ampere turns are needed than the conventional dipole for a given field. But for a superconducting magnet this problem is not too serious.

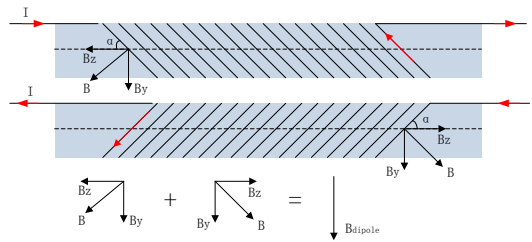

Fig.1 Filed patterns for superposition of two tilted solenoids

For this configuration, the conductor path of the coil is very like a series of ellipses connected together on the surface of a cylinder (see Fig.2). A three-dimensional curve can be used to define the current flow path expressing as (supposing the solenoid axis is along z direction):

$$x = R\cos\theta \qquad (1)$$

$$y = R\sin\theta \qquad (2)$$

$$z = R\sin\theta\cot\alpha + \frac{\theta h}{2\pi} \tag{3}$$

where R is the cylinder radius, θ is the azimuth angle and h is the winding pitch. So we need several cylindrical tubes with grooves having corresponding tilted helical geometry to place the conductor turns of each coil layers. These cylindrical tubes are also acted as coil former and the coil grooves can be machined by a computer-controlled milling machine easily. In view of this unique coil configuration having good prospects in various applications we have taken its R&D aiming at exploring the methods of design and construction for this type of coil. This work is also a pre-research for the superconducting accelerator magnets of HIAF project [7].

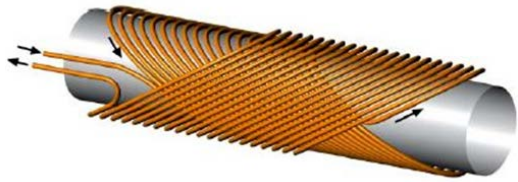

Fig.2 Schematic of two layers of tilted solenoids

## 2 Magnetic design

Because of much difference between the superconducting dipole magnet design based on titled solenoids and the regular superconducting magnet design we have done and having no experience, we start by developing a 2.5 T magnet prototype with 50 mm diameter and 500 mm long. For easily winding and less layers needing, we plan to employ a 7-stands cable to wrap the coils. The cable has a 2.2 mm diameter with insulation. In this section we will be devoted to the magnetic field design by using the software of RADIA and OPERA.

### 2.1 RADIA model

The key technology of the coil model creation is to define each turn position according to the above expressions. It is convenient and easy to make the modeling and analysis of this type of coils with RADIA by using line current model (regarding an actual conductor as a line through the center and its current is approximately concentrated in the line). Fig. 3 shows the line current coil model of four layers of tilted solenoids with an inclination angle of 30 degree [8]. The design parameters are listed in table 1.

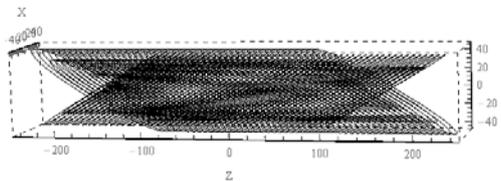

Fig.3 Tilted solenoids model by RADIA

Table 1 Design parameters of the tilted solenoids

| Parameter | Value |
| --- | --- |
| Coil aperture | 50 mm |
| Inclination angle | 30 deg |
| Winding pitch | 6.6 mm |
| Turns of each coil layer | 50 |

| Operating current | 3708 A |
| --- | --- |
| Total inductance | 1.1 mH |

The dipole field along the beam direction from -300 mm to 300 mm is plotted in Fig.4. We can find that the field decreases at about 100 mm away from the center in which the solenoids are superposed (see Fig.3). Thus this type of magnet can have a longer end than iron dominated magnet depending on the tilted angle. Fig.5 plots the fields distribute on a disk with 40 mm diameter at the magnet longitudinal center. Due to the different radius of each coil layer and incomplete symmetry of each coil itself, the field is not symmetry. The field homogeneity obtained on a 40 mm diameter circle at the cross-section is only $\pm 3 \times 10^{-4}$. However it has almost no influence on the integral field homogeneity (see Fig.6). Table 2 lists the multipole content at 20 mm reference radius at the longitudinal center.

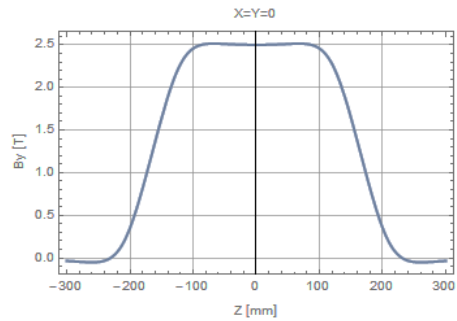

Fig.4 Dipole fields along the beam direction

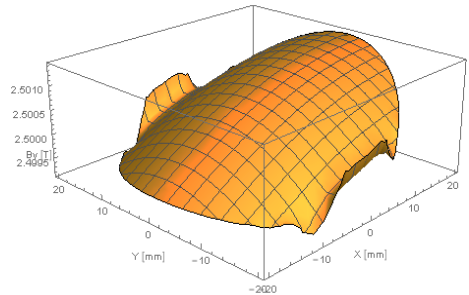

Fig.5 Dipole fields distribute on a 40 mm diameter disk

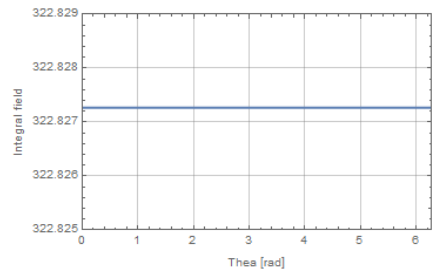

Fig.6 Integral field at different lines

Table 2 Multipole field (Gauss) at 20 mm reference radius at the longitudinal center

| Multipole order | Normal | Skew |
| --- | --- | --- |
| 0 | 25014.8 | |
| 1 | 1.13 | 0.5945 |
| 2 | 0.2456 | 0.6976 |
| 3 | 1.015 | -0.0308 |
| 4 | 0.2694 | 0.3588 |
| 5 | 0.4369 | 1.243 |

2.2 OPERA model

In OPREA we create the coils model by using a series of the 8-node brick conductors which is shown in Fig.7 [9]. Each of the conductors has a section of 1.95mm×1.95mm equaling the cross-section area of the 7-strands cable. Fig.8 shows the transverse field along the axial direction from -300 mm to 300 mm. Fig.9 plots a contour map of the transverse field on the 20 mm diameter at the longitudinal center. We can see that the field distribution simulated by OPERA is consistent with the result by RADIA.

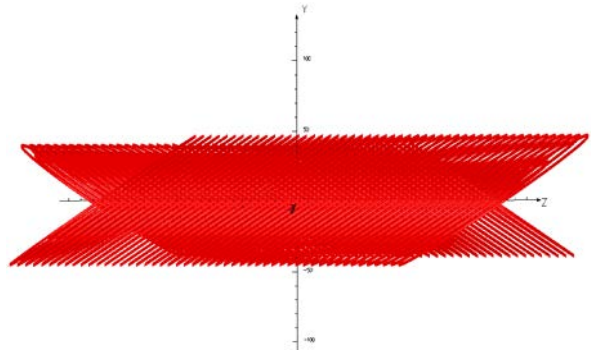

Fig.7 Prototype coil model by OPERA

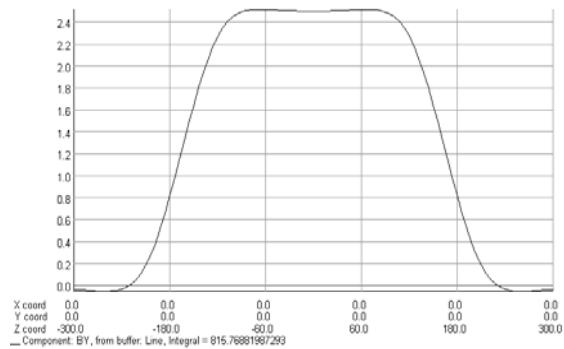

Fig.8 The transverse field along the axial direction

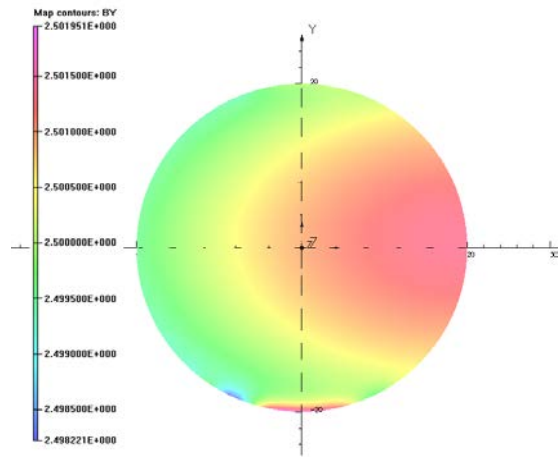

Fig.9 Contour map of transverse field on a disk

## 3 Construction of the prototype

The 2.5T magnet prototype consists of four coil layers, each of which needs to be wound in the grooves with tilted helical geometry on the G10 tubes. These grooves are machined by using a computer-controlled milling machine and one of the grooved tubes is shown in Fig.10. After the four coil layers are completed separately by placing the 7-strands round cable in the pre-machined grooves (see Fig.11), we start the assembly of the layers.

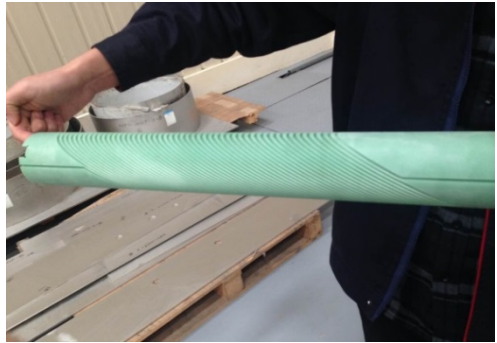

Fig.10 G10 tube with tilted helical geometry grooves

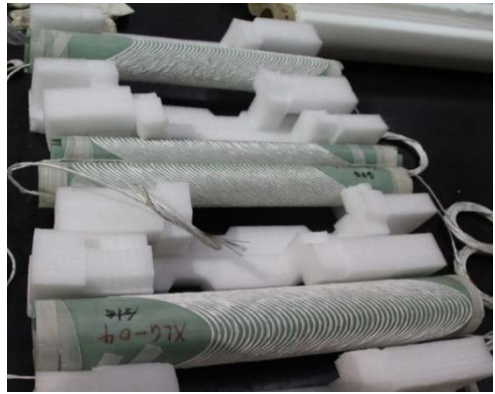

Fig.11 Four completed coil layers

A steal stainless bore tube is used to support the coil layers. We firstly put the innermost coil layer on the bore tube fixing the position through the end plate. Then the last three coil layers are installed one by one with 0.6 mm clearance between each two adjacent layers. For magnet reinforcement two layers of aluminum alloy wire are wound around the outermost coil layer to provide radial pre-compression. Finally, the completed coil assembly is vacuum impregnated with CTD-101 epoxy and the finished magnet prototype is shown in Fig.12. We can see that the construction of the tilted solenoids magnet is very easy and convenient by the modular approach.

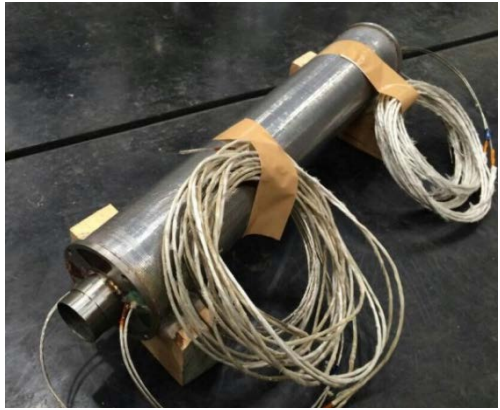

Fig.12 Photograph of the finished prototype

4 Conclusions

In this paper we have described the design of a dipole magnet with the novel coil configuration based on the tilted solenoids that has a good outlook for the future superconducting magnets. For research, we have fabricated a 2.5T superconducting prototype. The fields calculated by two software of RADIA and OPERA are well accorded. The field homogeneity is up to $\pm 3 \times 10^{-4}$ without optimization. Since a pair of 4 kA current leads for the prototype is under construction and will be probably completed in the second half of this year, the cryogenic testing will be made at that time.